\RequirePackage{lineno}  

\documentclass[aps,prl,preprint]{revtex4} 

\pagewiselinenumbers  

\usepackage{graphicx}
\usepackage{dcolumn}
\usepackage{bm}
\usepackage[dvips]{epsfig}

\begin{document}
\linenumbers

\title{Production of $\Lambda^{0}$, $\overline{\Lambda}^{0}$, $\Xi^{\pm}$, and $\Omega^{\pm}$
	Hyperons in $p\overline{p}$ Collisions at $\sqrt{s}=1.96$ TeV}
\date{\today}

\affiliation{Institute of Physics, Academia Sinica, Taipei, Taiwan 11529, Republic of China} 
\affiliation{Argonne National Laboratory, Argonne, Illinois 60439, USA} 
\affiliation{University of Athens, 157 71 Athens, Greece} 
\affiliation{Institut de Fisica d'Altes Energies, Universitat Autonoma de Barcelona, E-08193, Bellaterra (Barcelona), Spain} 
\affiliation{Baylor University, Waco, Texas 76798, USA} 
\affiliation{Istituto Nazionale di Fisica Nucleare Bologna, $^{bb}$University of Bologna, I-40127 Bologna, Italy} 
\affiliation{Brandeis University, Waltham, Massachusetts 02254, USA} 
\affiliation{University of California, Davis, Davis, California 95616, USA} 
\affiliation{University of California, Los Angeles, Los Angeles, California 90024, USA} 
\affiliation{Instituto de Fisica de Cantabria, CSIC-University of Cantabria, 39005 Santander, Spain} 
\affiliation{Carnegie Mellon University, Pittsburgh, Pennsylvania 15213, USA} 
\affiliation{Enrico Fermi Institute, University of Chicago, Chicago, Illinois 60637, USA}
\affiliation{Comenius University, 842 48 Bratislava, Slovakia; Institute of Experimental Physics, 040 01 Kosice, Slovakia} 
\affiliation{Joint Institute for Nuclear Research, RU-141980 Dubna, Russia} 
\affiliation{Duke University, Durham, North Carolina 27708, USA} 
\affiliation{Fermi National Accelerator Laboratory, Batavia, Illinois 60510, USA} 
\affiliation{University of Florida, Gainesville, Florida 32611, USA} 
\affiliation{Laboratori Nazionali di Frascati, Istituto Nazionale di Fisica Nucleare, I-00044 Frascati, Italy} 
\affiliation{University of Geneva, CH-1211 Geneva 4, Switzerland} 
\affiliation{Glasgow University, Glasgow G12 8QQ, United Kingdom} 
\affiliation{Harvard University, Cambridge, Massachusetts 02138, USA} 
\affiliation{Division of High Energy Physics, Department of Physics, University of Helsinki and Helsinki Institute of Physics, FIN-00014, Helsinki, Finland} 
\affiliation{University of Illinois, Urbana, Illinois 61801, USA} 
\affiliation{The Johns Hopkins University, Baltimore, Maryland 21218, USA} 
\affiliation{Institut f\"{u}r Experimentelle Kernphysik, Karlsruhe Institute of Technology, D-76131 Karlsruhe, Germany} 
\affiliation{Center for High Energy Physics: Kyungpook National University, Daegu 702-701, Korea; Seoul National University, Seoul 151-742, Korea; Sungkyunkwan University, Suwon 440-746, Korea; Korea Institute of Science and Technology Information, Daejeon 305-806, Korea; Chonnam National University, Gwangju 500-757, Korea; Chonbuk National University, Jeonju 561-756, Korea} 
\affiliation{Ernest Orlando Lawrence Berkeley National Laboratory, Berkeley, California 94720, USA} 
\affiliation{University of Liverpool, Liverpool L69 7ZE, United Kingdom} 
\affiliation{University College London, London WC1E 6BT, United Kingdom} 
\affiliation{Centro de Investigaciones Energeticas Medioambientales y Tecnologicas, E-28040 Madrid, Spain} 
\affiliation{Massachusetts Institute of Technology, Cambridge, Massachusetts 02139, USA} 
\affiliation{Institute of Particle Physics: McGill University, Montr\'{e}al, Qu\'{e}bec, Canada H3A~2T8; Simon Fraser University, Burnaby, British Columbia, Canada V5A~1S6; University of Toronto, Toronto, Ontario, Canada M5S~1A7; and TRIUMF, Vancouver, British Columbia, Canada V6T~2A3} 
\affiliation{University of Michigan, Ann Arbor, Michigan 48109, USA} 
\affiliation{Michigan State University, East Lansing, Michigan 48824, USA}
\affiliation{Institution for Theoretical and Experimental Physics, ITEP, Moscow 117259, Russia}
\affiliation{University of New Mexico, Albuquerque, New Mexico 87131, USA} 
\affiliation{Northwestern University, Evanston, Illinois 60208, USA} 
\affiliation{The Ohio State University, Columbus, Ohio 43210, USA} 
\affiliation{Okayama University, Okayama 700-8530, Japan} 
\affiliation{Osaka City University, Osaka 588, Japan} 
\affiliation{University of Oxford, Oxford OX1 3RH, United Kingdom} 
\affiliation{Istituto Nazionale di Fisica Nucleare, Sezione di Padova-Trento, $^{cc}$University of Padova, I-35131 Padova, Italy} 
\affiliation{LPNHE, Universite Pierre et Marie Curie/IN2P3-CNRS, UMR7585, Paris, F-75252 France} 
\affiliation{University of Pennsylvania, Philadelphia, Pennsylvania 19104, USA}
\affiliation{Istituto Nazionale di Fisica Nucleare Pisa, $^{dd}$University of Pisa, $^{ee}$University of Siena and $^{ff}$Scuola Normale Superiore, I-56127 Pisa, Italy} 
\affiliation{University of Pittsburgh, Pittsburgh, Pennsylvania 15260, USA} 
\affiliation{Purdue University, West Lafayette, Indiana 47907, USA} 
\affiliation{University of Rochester, Rochester, New York 14627, USA} 
\affiliation{The Rockefeller University, New York, New York 10065, USA} 
\affiliation{Istituto Nazionale di Fisica Nucleare, Sezione di Roma 1, $^{gg}$Sapienza Universit\`{a} di Roma, I-00185 Roma, Italy} 

\affiliation{Rutgers University, Piscataway, New Jersey 08855, USA} 
\affiliation{Texas A\&M University, College Station, Texas 77843, USA} 
\affiliation{Istituto Nazionale di Fisica Nucleare Trieste/Udine, I-34100 Trieste, $^{hh}$University of Trieste/Udine, I-33100 Udine, Italy} 
\affiliation{University of Tsukuba, Tsukuba, Ibaraki 305, Japan} 
\affiliation{Tufts University, Medford, Massachusetts 02155, USA} 
\affiliation{Waseda University, Tokyo 169, Japan} 
\affiliation{Wayne State University, Detroit, Michigan 48201, USA} 
\affiliation{University of Wisconsin, Madison, Wisconsin 53706, USA} 
\affiliation{Yale University, New Haven, Connecticut 06520, USA} 
\author{T.~Aaltonen}
\affiliation{Division of High Energy Physics, Department of Physics, University of Helsinki and Helsinki Institute of Physics, FIN-00014, Helsinki, Finland}
\author{B.~\'{A}lvarez~Gonz\'{a}lez$^v$}
\affiliation{Instituto de Fisica de Cantabria, CSIC-University of Cantabria, 39005 Santander, Spain}
\author{S.~Amerio}
\affiliation{Istituto Nazionale di Fisica Nucleare, Sezione di Padova-Trento, $^{cc}$University of Padova, I-35131 Padova, Italy} 

\author{D.~Amidei}
\affiliation{University of Michigan, Ann Arbor, Michigan 48109, USA}
\author{A.~Anastassov}
\affiliation{Northwestern University, Evanston, Illinois 60208, USA}
\author{A.~Annovi}
\affiliation{Laboratori Nazionali di Frascati, Istituto Nazionale di Fisica Nucleare, I-00044 Frascati, Italy}
\author{J.~Antos}
\affiliation{Comenius University, 842 48 Bratislava, Slovakia; Institute of Experimental Physics, 040 01 Kosice, Slovakia}
\author{G.~Apollinari}
\affiliation{Fermi National Accelerator Laboratory, Batavia, Illinois 60510, USA}
\author{J.A.~Appel}
\affiliation{Fermi National Accelerator Laboratory, Batavia, Illinois 60510, USA}
\author{A.~Apresyan}
\affiliation{Purdue University, West Lafayette, Indiana 47907, USA}
% ...
\collaboration{CDF Collaboration\footnote{With visitors from $^a$University of Massachusetts Amherst, Amherst, Massachusetts 01003,
$^b$Istituto Nazionale di Fisica Nucleare, Sezione di Cagliari, 09042 Monserrato (Cagliari), Italy,
$^c$University of California Irvine, Irvine, CA  92697, 
$^d$University of California Santa Barbara, Santa Barbara, CA 93106
$^e$University of California Santa Cruz, Santa Cruz, CA  95064,
$^f$CERN,CH-1211 Geneva, Switzerland,
$^g$Cornell University, Ithaca, NY  14853, 
$^h$University of Cyprus, Nicosia CY-1678, Cyprus, 
$^i$University College Dublin, Dublin 4, Ireland,
$^j$University of Fukui, Fukui City, Fukui Prefecture, Japan 910-0017,
$^k$Universidad Iberoamericana, Mexico D.F., Mexico,
$^l$Iowa State University, Ames, IA  50011,
$^m$University of Iowa, Iowa City, IA  52242,
$^n$Kinki University, Higashi-Osaka City, Japan 577-8502,
$^o$Kansas State University, Manhattan, KS 66506,
$^p$University of Manchester, Manchester M13 9PL, England,
$^q$Queen Mary, University of London, London, E1 4NS, England,
$^r$Muons, Inc., Batavia, IL 60510,
$^s$Nagasaki Institute of Applied Science, Nagasaki, Japan, 
$^t$National Research Nuclear University, Moscow, Russia,
$^u$University of Notre Dame, Notre Dame, IN 46556,
$^v$Universidad de Oviedo, E-33007 Oviedo, Spain, 
$^w$Texas Tech University, Lubbock, TX  79609, 
$^x$IFIC(CSIC-Universitat de Valencia), 56071 Valencia, Spain,
$^y$Universidad Tecnica Federico Santa Maria, 110v Valparaiso, Chile,
$^z$University of Virginia, Charlottesville, VA  22906,
$^{aa}$Yarmouk University, Irbid 211-63, Jordan,
$^{ii}$On leave from J.~Stefan Institute, Ljubljana, Slovenia, 
}}
\noaffiliation


\begin{abstract}
We report a set of measurements of inclusive invariant $p_T$ differential 
cross sections of $\Lambda^{0}$, $\overline{\Lambda}^{0}$, $\Xi^{\pm}$, and $\Omega^{\pm}$ hyperons 
reconstructed in the central region with pseudorapidity $|\eta|<1$ and  $p_T$ up to 10 GeV$/c$. 
Events are collected with a minimum-bias trigger in $p \bar p$ collisions at a 
center-of-mass energy of 1.96 TeV using the CDF II detector at  the Tevatron Collider.
As $p_T$ increases, the slopes of the differential cross sections are similar not only to each other
but also to those of mesons, which could indicate a universality of the particle production in $p_T$
The invariant differential cross sections are also presented for different
charged-particle multiplicity intervals.
\end{abstract}


\maketitle

 Ever since their discovery in cosmic ray interactions~\cite{c:a}, 
 particles containing strange quarks have been extensively studied at
 particle colliders ($e^+e^-$~\cite{c:aa}, $ep$~\cite{c:ab},
 $p\bar{p}$~\cite{c:ac,c:gg} and $pp$~\cite{c:ad}). The process by which hadrons
 in general are produced from interactions is an unsolved 
 problem in the standard model, and a detailed analysis of production 
 properties of particles with different quark flavors and numbers of quarks could 
 pave the way to  understanding  the process from first principles.
 The data on strange particle production  can also be used to 
 refine phenomenological models and set parameters, such as the strange quark 
 suppression constant in  event generators, 
 which have become an integral part of any data analysis. 
 Interest in particles containing strange quarks increased with the introduction of
 the quark-gluon plasma (QGP). Formation of quark-gluon plasma in a
 collision could manifest itself as an enhanced production of strange
 particles such as kaons and hyperons~\cite{c:ae}. To isolate QGP 
 signatures in heavy-ion collision data, understanding the particle 
 production properties from simple nucleon interactions is necessary.

 There are ample data on the production of particles with one strange quark, 
 but very little available on particles with two or more~\cite{c:b,c:c}.
 Previous studies of hyperons from colliders such as RHIC~\cite{c:d},
 $Sp\overline{p}S$~\cite{c:e}, and the Tevatron~\cite{c:f,c:g} were limited
 by low sample statistics and the limited accessible range of hyperon momentum component
 transverse to the beam direction ($p_T$).
 In this Letter, we report on a study of the hyperons $\Lambda^0$
 (quark content $uds$), $\Xi^-$ ($dss$), and $\Omega^-$ ($sss$) and their
 corresponding antiparticles ($\overline{\Lambda}^0$, $\Xi^+$, and $\Omega^+$).
 For these hyperons, the inclusive invariant $p_T$ differential cross 
 sections are measured up to $p_T$ of 10 GeV/$c$, based on $\sim100$ million 
 minimum-bias events collected with the CDF II detector. The
 measurements reported  here are the current best from any hadron 
 collider experiment in terms of statistics and $p_T$ range.

 The CDF II detector is described in detail elsewhere~\cite{c:h}. 
 The components most relevant to this analysis are those that comprise the 
 tracking system, which is within a uniform axial magnetic field of $1.4$T. 
 The inner tracking volume is composed of a system of eight layers of 
 silicon microstrip detectors ranging in radius from 1.5 to
 28.0 cm~\cite{Sill:2000zz} in the pseudorapidity region $|\eta|<2$~\cite{fn:a}.
 The remainder of the tracking volume is occupied by the Central Outer Tracker
 (COT). The COT is a cylindrical drift chamber containing 96 sense-wire layers 
 grouped in eight alternating superlayers of axial and stereo
 wires~\cite{Affolder:2003ep}. Its active volume covers 40 to 140 cm in radius
 and $|z|<155$ cm. The transverse-momentum resolution of tracks reconstructed 
 using COT hits is $\sigma(p_T)/p_T^2 \sim0.0017$/(GeV/$c$). 

 Events for this analysis are collected with a ``minimum-bias'' (MB) trigger,
 which selects beam crossings with at least one
 $p\bar{p}$ interaction by requiring a timing coincidence for signals 
 in both  forward and backward gas Cherenkov counters ~\cite{Acosta:clc}
 covering the regions $3.7 < |\eta| < 4.7$. 
 The MB trigger is rate-limited to keep the
 final trigger output at 1 Hz. Primary event vertices are identified by the 
 convergence of reconstructed tracks along the beam axis. Events are accepted that
 contain a reconstructed vertex in the fiducial region $|z_{vtx}| \le 60$ cm
 centered around the nominal CDF origin ($z=0$).
 When an event has more than one vertex, the highest quality vertex, 
 usually the one with the most associated tracks, is selected and it is 
 required that there be no other vertices within $\pm5$ cm of this vertex.
 This selection introduces a bias toward high multiplicity events as the 
 instantaneous luminosity increases. To combine events collected at different 
 average instantaneous luminosities, we determine a per-event weight as a
 function of the charged-track multiplicity $N_{ch}$ in order 
 to match the multiplicity distribution of a data sample 
 where the average number of interactions is less than 0.3 per bunch
 crossing. For the $N_{ch}$ calculation, tracks are required to have a high 
 track-fit quality with $\chi^2$ per degree-of-freedom ($\chi^2/dof$) less
 than $2.5$, and more than five hits in at least two axial and two stereo
 COT segments. It is further required that tracks satisfy $|\eta|<1$, 
 impact parameter  $d_0$~\cite{fn:b}
 less than 0.25 cm, the distance along the $z$-axis ($\delta Z_0$)between the
 event vertex and the track position at the point of closest approach
 to the vertex in the $r-\phi$ plane be less than $2$ cm, and $p_T > 0.3$ GeV$/c$.
 The $p_T$ selection is to minimize the inefficiency of the track-finding algorithm 
 for low momentum tracks. 

 We search for $\Lambda^0 \to p\pi^-$ decays using tracks with opposite-sign charge and 
 $p_T>0.325$ GeV/$c$ that satisfy the $\chi^2/dof$ and COT segment requirements.
 In this Letter, any reference to a specific hyperon state implies the 
 antiparticle state as well.
 For each two-track combination we calculate their intersection coordinate in the
 $r-\phi$ plane. Once this intersection point, referred to as the secondary
 vertex, is found, the $z$-coordinate of each track ($Z_{1}$ and $Z_{2}$) is
 calculated at that point. If the distance $|Z_1 - Z_2|$ is less than 1.5 cm,
 the tracks are considered to originate from a $\Lambda^0$ candidate decay. The pair
 is traced  back to the vertex and we require $\delta Z_0$ be less than $2$ cm, 
 and the $d_0$ be less than $0.25$ cm. To reduce backgrounds further, we require the 
 $\Lambda^0$ decay length $L_{\Lambda^0}$, the distance in the $r-\phi$ plane 
 between the primary and secondary vertices, to be greater than $2.5$ cm and
 less than $50$ cm.

 The invariant mass $M_{p\pi}$ of the two-track system is calculated by 
 attributing the proton mass to the track with the higher $p_T$, as preferentially expected 
 by the kinematics of a $\Lambda^0$ decay. Figure~\ref{f:a} shows the invariant
 mass for $\Lambda^0$ candidates with $|\eta|<1$. This distribution is divided into
 23 $p_T$ intervals~\cite{fn:bb} and the number of $\Lambda^0$ in each $p_T$ interval
 is determined by fitting the invariant mass distributions using a Gaussian
 function with three parameters for the signal and a third-order polynomial for the 
 underlying combinatorial background.  The data in the mass range $1.10-1.16$ GeV/$c^2$ are fitted. 
 The polynomial fit to the background is subtracted bin-by-bin from the
 data entries in the $\Lambda^0$ mass window ($1.111 - 1.121$ GeV/$c^2$) to obtain 
 the number of  $\Lambda^0$ hyperons. This number is divided by the acceptance to
 obtain the invariant $p_T$ differential cross section as described later. 

 The fitting procedure is one source of systematic uncertainty. This uncertainty 
 is estimated by separately varying the mass range of the fit, the functional form for
 the signal to a double Gaussian, and the background modeling function to a second-order polynomial. 
 The number of $\Lambda^0$ is recalculated in all $p_T$ intervals for each variation. 
 The systematic uncertainty is determined as the sum in quadrature 
 of the fractional change in the number of $\Lambda^0$
 from each modified fit. It decreases from ~$\pm 10\%$ at the lowest 
 $p_T$ (1.2 GeV/$c$)  to less than $\pm 5\%$ for  $p_T > 1.75$ GeV$/c$.
 
 The cascade reconstruction  decay mode is
 $\Xi^{-}\rightarrow\Lambda^0\pi^{-}\rightarrow(p\pi^{-})\pi^{-}$.
 The previously reconstructed $\Lambda^0$ candidates are used, but without 
 the $d_0$ and $\delta Z_{0}$ requirements. We select $\Lambda^0$ candidates 
 in the $\Lambda^0$ mass window and calculate the coordinate of the intersection 
 point in the $r-\phi$ plane between the $\Lambda^0$ candidate and a third
 track. The $z$-axis coordinates at this
 point are calculated for the third track ($Z_3$) and the $\Lambda^0$ candidate
 ($Z_4$). The three-track system is considered a $\Xi^-$ candidate decay if the 
 distance $|Z_3-Z_4|<1.5$ cm. We also require the decay length $L_{\Xi^-} > 1$ cm, and
 $(L_{\Lambda^0}-L_{\Xi^-}) >1$ cm.
 Finally, it is required that the $d_0$ and $\delta Z_0$ of the $\Xi^-$ candidate be less 
 than $0.25$ cm and $2$ cm respectively.

 The invariant mass $M_{\Lambda^0\pi}$ is calculated by fixing the mass of the 
 $\Lambda^0$ candidate to 1.1157 GeV/$c^{2}$~\cite{c:o} and assigning the
 pion mass to the third track. Figure~\ref{f:a} shows the invariant mass 
 for $\Xi^-$ candidates with $|\eta|<1$.
  
 The $\Xi^-$ candidates are divided 
 into 17 $p_T$ intervals and the number of $\Xi^-$ in each interval is determined 
 by fitting the corresponding $M_{\Lambda^0\pi}$ invariant mass distribution using 
 a Gaussian function for the signal and a third-order polynomial for the 
 background. The fitted background is then subtracted bin-by-bin from the 
 data entries in the signal region (1.31 to 1.33 GeV/$c^2$) 
 to obtain the $\Xi^-$ yield in every $p_T$ 
 interval. The systematic uncertainty of the fit procedure is estimated the same 
 way as for the $\Lambda^0$ and is found to change by no more than $\pm 5$\% 
 in all  $p_T$ intervals. 
 
  To reconstruct $\Omega^-$ decays we follow the same procedure as for the 
  $\Xi^-$ and apply the same selection criteria except that the third track 
  is assigned the kaon mass. The search decay mode is 
  $\Omega^{-}\rightarrow\Lambda^0K^{-}\rightarrow(p\pi^{-}) K^{-}$.
  Because of the larger background, the procedure to extract the $\Omega^{-}$ signal yield
  is slightly different from that in the previous cases. 
  Track pairs with $M_{p\pi^-}$ in the mass ranges $1.095-1.105$ and
  $1.127-1.137$ GeV/$c^2$ are combined with the third track to obtain 
  the invariant mass distribution of the combinatorial background. This
  distribution is subtracted from the $M_{\Lambda^0 K^-}$ distribution after
  normalizing to the number of events in the mass window
  $1.69 < M_{\Lambda^0K^-} < 1.74$ GeV/$c^2$. The background subtracted
  $M_{\Lambda^0K^-}$ invariant mass distribution is shown in Fig.~\ref{f:a}. 

  The distribution is divided into 10 $p_T$ intervals, and we use
  the method described above to extract the $\Omega^-$ signal from
  the corresponding invariant mass distributions in each $p_T$ interval within the
  mass window 1.66 to 1.68 GeV/$c^2$.
  The systematic  uncertainty due to the fitting procedure is also calculated 
  in a similar manner as $\Xi^-$, with the exception of using a double Gaussian variation 
  because of low $\Omega^-$ statistics. The overall uncertainties are about $\pm 10$\% 
  for all $p_T$ intervals.

 The geometric and kinematic acceptance is estimated with Monte Carlo (MC) 
 simulations. The MC data of a resonance state
 are generated with fixed $p_T$ corresponding to 14 points~\cite{fn:bb} ranging 
 from 0.75 to 10 GeV/$c$ and flat in rapidity $|y|<2$. A generated 
 resonance is combined with either one or four non-diffractive inelastic MB events generated
 with the {\sc pythia} {\cite{c:k} generator. Although the average number 
 of interactions in our data sample is a little less than two,
 the default acceptance is calculated from the MC sample with four MB events
 and the difference of the acceptance values between the two samples is one of 
 our systematic uncertainties. Based on a study with tracks from $K_{S}^{0}$ decay,
 the sample with four MB  events reproduces the low $p_T$ tracking efficiency  
 in data well within the systematic uncertainty

 The detector response to particles produced in the simulation is modeled
 with the CDF II detector simulation that in turn is based on the {\sc geant-3}
 MC program~\cite{GEANT}. Simulated events are processed and selected
 with the same analysis code used for the data. The acceptance
 is defined as the ratio of the  number of reconstructed resonances with
 the input $p_T$ over the generated number, including the branching ratio. 
 Acceptance values are calculated separately for the particles and their
 corresponding antiparticles and the average of the two is used as the 
 default value, since the acceptances for the two states are similar. 
 The acceptance values obtained for the 14 $p_T$ points are fitted with a fourth
 order polynomial function and the fitted curve is used to correct the numbers
 of each hyperon state in the data. 
 
 The modeling of the MB events overlapping with the examined resonance and 
 the selection criteria applied contribute as a systematic uncertainty
 to the acceptance calculation. The contribution from the former has  already been mentioned. 
 Acceptance uncertainties due to the selection criteria are studied by changing
 the selection values of the variables used to reconstruct the resonances. The variables 
 examined are $p_T$, $|Z_1 - Z_2|$, $|Z_3 - Z_4|$, $\delta Z_0$, $d_0$ and the decay lengths. For
 each variable other than $p_T$, two values around the default value are typically chosen.
 One value is such that it has little effect on the signal, and the other reduces the 
 signal by $\sim$20 to 30$\%$. The default $p_T$ selection value is 0.325 GeV$/c$, and 
 it is changed to 0.3 GeV$/c$ and to 0.35 GeV$/c$. 

 For each considered variation, a new acceptance curve and the number of resonances
 as a function of $p_T$
 are obtained, and the percentage change between the new $p_T$ distribution and the 
 one with the default selection requirements is taken as the uncertainty 
 in the acceptance for the specific $p_T$ interval. The square root of the 
 quadratic sum of the uncertainties from each variation is taken as the 
 total conservative uncertainty on the acceptance in a given $p_T$ bin.
 The systematic uncertainty associated with the $\Omega^{-}$ hyperon
 acceptance is derived from the $\Xi^{-}$ uncertainty estimate since the
 reconstruction follows the same criteria. This acceptance uncertainty is 
 added quadratically to the systematic uncertainty due to the
 fitting procedure, described earlier, to give the total systematic uncertainty.

 For the $\Lambda^0$ case, the acceptance uncertainty decreases from about
 $25\%$ at $p_T \sim 1$ GeV/$c$ to $10\%$ at $p_T \sim 2$ GeV/$c$ and then
 rises again slowly to $15\%$ for $p_T>7$ GeV/$c$.
 The corresponding acceptance uncertainty for the $\Xi^-$ ($\Omega^-$) case decreases
 from about 15\% (20\%) at $p_T\sim2$ GeV$/c$ to 10\% (15\%)  for $p_T > 4$ GeV$/c$.

 The inclusive invariant $p_T$ differential cross section for each hyperon  
 resonance is calculated as $Ed^3\sigma/dp^3$ = $(\sigma_{mb}/N_{event}) d^3N/A p_T dp_T dy d\phi$ 
 = $(\sigma_{mb}/2\pi N_{event}) \Delta N/A p_T  \Delta p_T \Delta y $ where $\sigma_{mb}$ is our 
 MB trigger cross section,  $N_{event}$ is the number of weighted events, $\Delta N$ is 
 the number of hyperons observed in each $p_T$ interval ($\Delta p_T$) after 
 background subtraction, $A$ is the acceptance in the specific $p_T$ interval,
 and $\Delta y$ is the rapidity range used in the acceptance calculation (-2 to 2).

 Figure~\ref{f:b} shows the results for the  
 $p_T$ differential cross section for the three hyperon resonances. The uncertainties 
 shown for each data point include the statistical and all systematic
 uncertainties described above, except the one associated with $\sigma_{mb}$~\cite{c:l}. 

 The $p_T$ differential cross section is modeled by
 a power law function, $A(p_0)^n/(p_T +p_0)^n$, for $p_T > 2$ GeV$/c$.
 In order to compare with the previous CDF $K_{S}^{0}$ result~\cite{c:m,c:gg},
 $p_0$ is fixed at 1.3 GeV$/c$, and the results are shown
 in Tab. \ref{t.a}.  The data below $p_T \sim 2$ GeV$/c$ cannot be
 described well by the power law function even if $p_0$ is allowed to
 float.  For this region, the data are better described by an exponential
 function, $B\exp [ -b \cdot p_T ]$. The results of this fit are
 shown in Tab.~\ref{t.b}, and the slope $b$ of $\Lambda$ is consistent with
 previous measurements \cite{c:f,c:g}. 
 The $b$ values depend on the range of the fit but are about
 two, which corresponds to an average $p_T$ of 1 GeV/$c$ under the
 assumption that the fit can be extrapolated down to $p_{T}=0$ GeV/$c$.

\begin{table}
\begin{center}
\caption{\label{t.a} The results of power law function fits to the 
        $p_T$ differential cross sections described in the text and shown in Fig.~\ref{f:b} 
	for $p_T>2$ GeV$/c$.  The parameter $p_0$ is fixed to $1.3$ GeV$/c$ in 
	all fits. The $K_S^0$ values are from Ref.~\cite{c:m} at $\sqrt{s}=1.8$ 
	TeV. The uncertainties shown do not include the MB cross section 
	uncertainty~\cite{c:l}.  The last line of the table gives the $\chi^2$
	per degree-of-freedom of the fit to data.}
\begin{tabular}{ccccc}
\hline\hline
Parameter (units) & $K_S^0$~\cite{c:m} & $\Lambda^0$ & $\Xi^{\pm}$ & $\Omega^{\pm}$ \\ 
\hline
$A$ (mb/GeV$^{2}c^{3}$) & $45\pm9$    & $210\pm25$    & $14.9\pm2.5$  & $1.50\pm0.75$ \\
$p_0$ (GeV$/c$)		& $1.3$	      & $1.3$	      & $1.3$	      & $1.3$	      \\
$n$ 			& $7.7\pm0.2$ & $8.81\pm0.08$ & $8.26\pm0.12$ & $8.06\pm0.34$ \\ 
$\chi^2$/\textit{dof}   & $8.1/11$    & $5.7/15$      & $15.8/15$     & $10.5/7$      \\ 
\hline\hline                                                       
\end{tabular}
\end{center}
\end{table}

\begin{table}
\begin{center}
\caption{\label{t.b} The results of exponential function fits to the 
        $p_T$ differential cross sections shown in Fig.~\ref{f:b} 
	for the $p_T$ ranges given in the second row.  The uncertainties shown
        do not include the MB cross section uncertainty~\cite{c:l}.
	The last line of the table gives the $\chi^2$ per degree-of-freedom of 
	the fit to data.}
\begin{tabular}{ccccc}\hline\hline
Parameter (units) & $\Lambda^0$ & $\Lambda^0$ & $\Xi^{\pm}$ & $\Omega^{\pm}$ \\
\hline
$p_T$ range (GeV/$c$) & [1.2, 2.5] & [1.2, 4] & [1.5, 4] & [2, 4] \\
$B$ (mb/GeV$^{2}c^{3}$) & $4.68\pm1.04$ & $3.16\pm0.35$ & $0.16\pm0.04$ & $0.024\pm0.011$ \\ 
$b$ (GeV$^{-1}c$)       & $2.30\pm0.12$ & $2.10\pm0.04$ & $1.75\pm0.08$  & $1.80\pm0.19$   \\
$\chi^2/dof$            & $1.0/7$       & $7.2/12$      & $4.0/8$        & $6.3/3$         \\ 
\hline \hline                                                       
\end{tabular}
\end{center}
\end{table}    


 The plots on the right side of Fig.~\ref{f:b} show the ratio of the $p_T$ differential cross
 sections for $\Xi^-$ and $\Lambda^0$, and $\Omega^-$ and $\Lambda^0$.
 In the $\Xi^-/\Lambda^0$ ratio there is a rise at low $p_T$,
 and the ratio reaches a plateau at $p_T > 4$ GeV$/c$.  It should be noted that
 the $\Lambda^0$ cross section also includes $\Lambda^0$ production
 from the decay of other hyperon states
 ($\Sigma^{0}\rightarrow\Lambda^{0}\gamma$, $\Xi^{\pm}$, $\Xi^{0}$ and
 $\overline{\Xi}^{0}$).  Due to the short $\Sigma^{0}$ lifetime,
 $\Lambda^0$ from $\Sigma^{0}$ decays cannot be separated from direct
 $\Lambda^0$ production.
 Simulations of cascade decays indicate that $\sim50$\% of $\Lambda^0$
 from $\Xi$ decays will satisfy our $\Lambda^0$ selection criteria, with
 the fraction of $\Lambda^0$ fairly independent of $\Xi$ $p_T$.
 The ratio plots in Fig.~\ref{f:b} are fitted to a constant, and the
 value $0.17\pm0.01$ is obtained for $\Xi^-/\Lambda^0$ and
 $0.025\pm0.002$ for $\Omega^-/\Lambda^0$. 

 The plots in Fig.~\ref{f:b} clearly show that the cross sections
 depend on the number of strange quarks. 
 However, the similarity of the $n$ values in  Tab. \ref{t.a}, along with the measured
 value $n = 8.28 \pm 0.02$ for all charged particles [26], indicate that the slope
 of the $p_T$ differential cross sections are similar in the high $p_T$ region.
 This could be an indication of a universality in particle production as $p_T$ increases~\cite{c:pq}. 
 This is in contrast to the low $p_T$ region where 
 the slope exhibits a strong particle type dependence~\cite{c:pp}. 
 

 Figure~\ref{f:c} shows the $p_T$ differential cross sections for two
 charged-particle multiplicity regions, $N_{ch}< 10$ and
 $N_{ch}>24$. $N_{ch}=24~(10)$ corresponds to $dN/d\eta \sim 16~(7)$,
 corrected for  the track reconstruction efficiency and unreconstructed
 tracks with $p_T < 0.3$ GeV$/c$~\cite{c:pr}.  Due to the low $\Omega^-$ sample
 statistics, distributions are only shown for $\Lambda^0$ and $\Xi^-$.  
 We observe a correlation between high $p_T$ particles and high multiplicity events.
 This is a general characteristic independent of the particle types.

 In summary, the production properties of $\Lambda^0$, $\Xi^-$, and
 $\Omega^-$ hyperons reconstructed from minimum-bias events at
 $\sqrt{s} = 1.96$ TeV are studied.  The inclusive invariant $p_T$
 differential cross sections are well modeled by a power law function
 above 2 GeV/$c$ $p_T$. With fixed $p_{0}$, the fit parameter $n$
 decreases from $8.81\pm0.08$ ($\Lambda^0$) to  $8.06\pm0.34$ ($\Omega^-$). 
 The low $p_{T}$ regions are modeled by an exponential function.  The
 exponential slope, $b$, decreases by $\sim15\%$ from $\Lambda^0$ to $\Omega^-$.
 The cross section ratios $\Xi^-/\Lambda^0$ and $\Omega^-/\Lambda^0$ are
 presented as a function of $p_T$. Although the ratios exhibit a strong dependence on
 the number of strange quarks, the $n$ values of the hyperons, $K_S^0$ and all charged particles
 are within $\sim 10\%$ of each other. This could be an indication that the production process which
 determines  the $p_T$ of these particles depends little on the particle type as $p_T$ increases. 
 We also find the hyperon $p_T$ differential cross sections fall off faster
 with $p_T$ for low multiplicity events than for high multiplicity
 events.

\begin{figure*}[!hp]
{\psfig{file=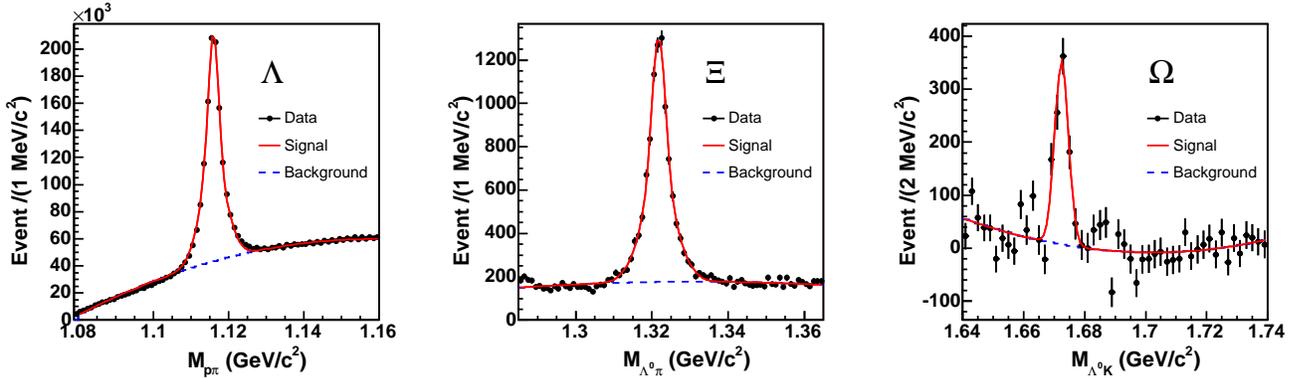,height=5.5cm}}
\caption{Reconstructed invariant mass distributions for $M_{p\pi}$ (left),
	$M_{\Lambda^0\pi}$ (center), and $M_{\Lambda^0 K}$ (right).  The
	background has been subtracted from the $M_{\Lambda^0 K}$
	distribution.  The solid lines are fitted curves, a
	third-degree polynomial for the background and either a double
	($M_{p\pi}$ and $M_{\Lambda^0\pi}$) or single ($M_{\Lambda^0 K}$)
	Gaussian function to model the peak. The widths reflect the tracking
        resolution and are consistent with the widths from MC simulation.}
\label{f:a}
\end{figure*}

\begin{figure*}[!hp]
{\psfig{file=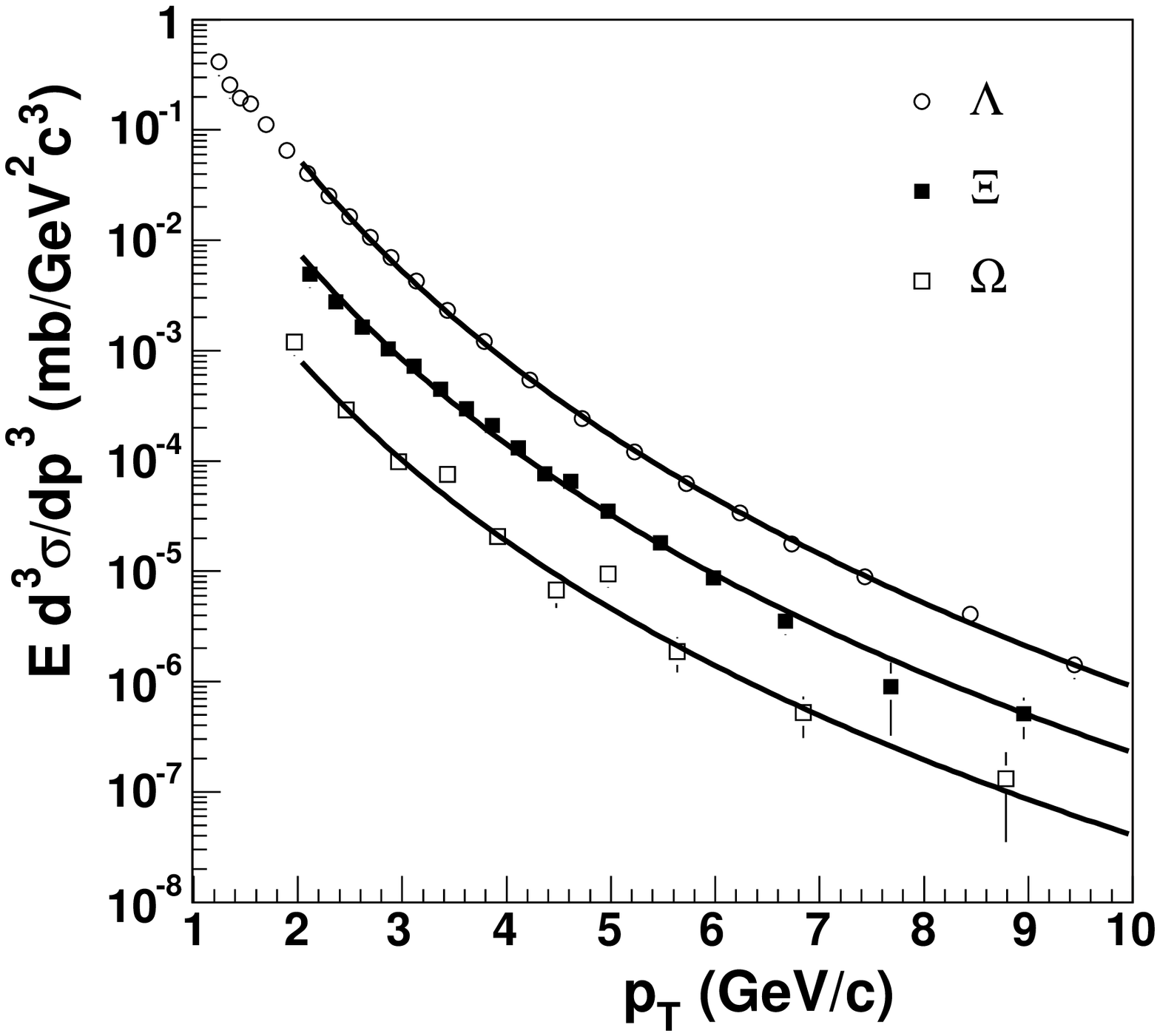,height=7cm}}
{\psfig{file=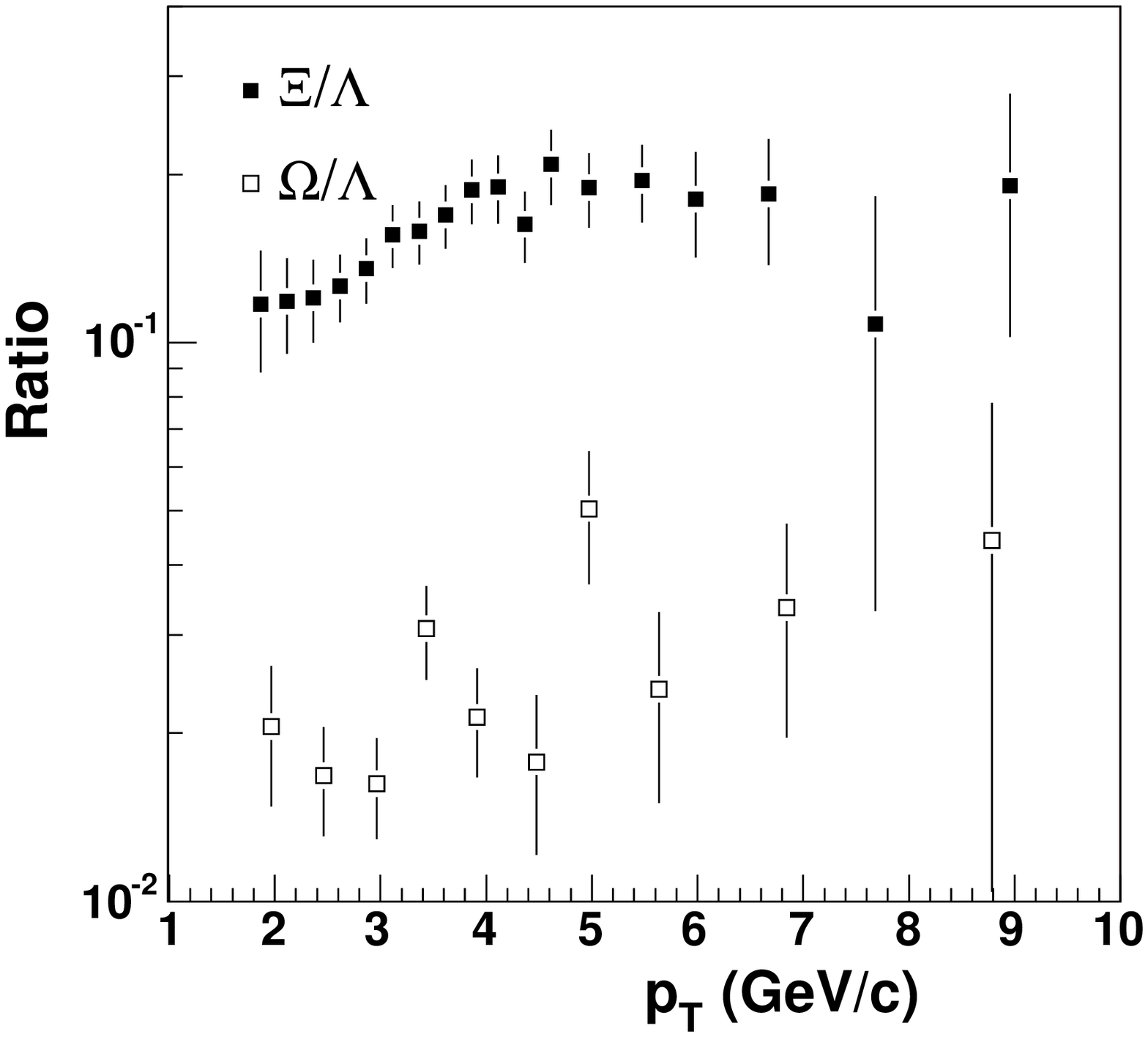,height=7cm}}
\caption{The $p_T$ differential cross sections for $\Lambda^0$, $\Xi^-$, and
	$\Omega^-$ within $|\eta|<1$ (left). The solid curves are from fits
	to a power law function, with the fitted parameters given in
	Tab. \ref{t.a}.  The ratios of
	$\Xi^-/\Lambda^0$ and $\Omega^-/\Lambda^0$ as a function of
	$p_T$ (right).}
\label{f:b}
\end{figure*}

\begin{figure*}[!hp]
{\psfig{file=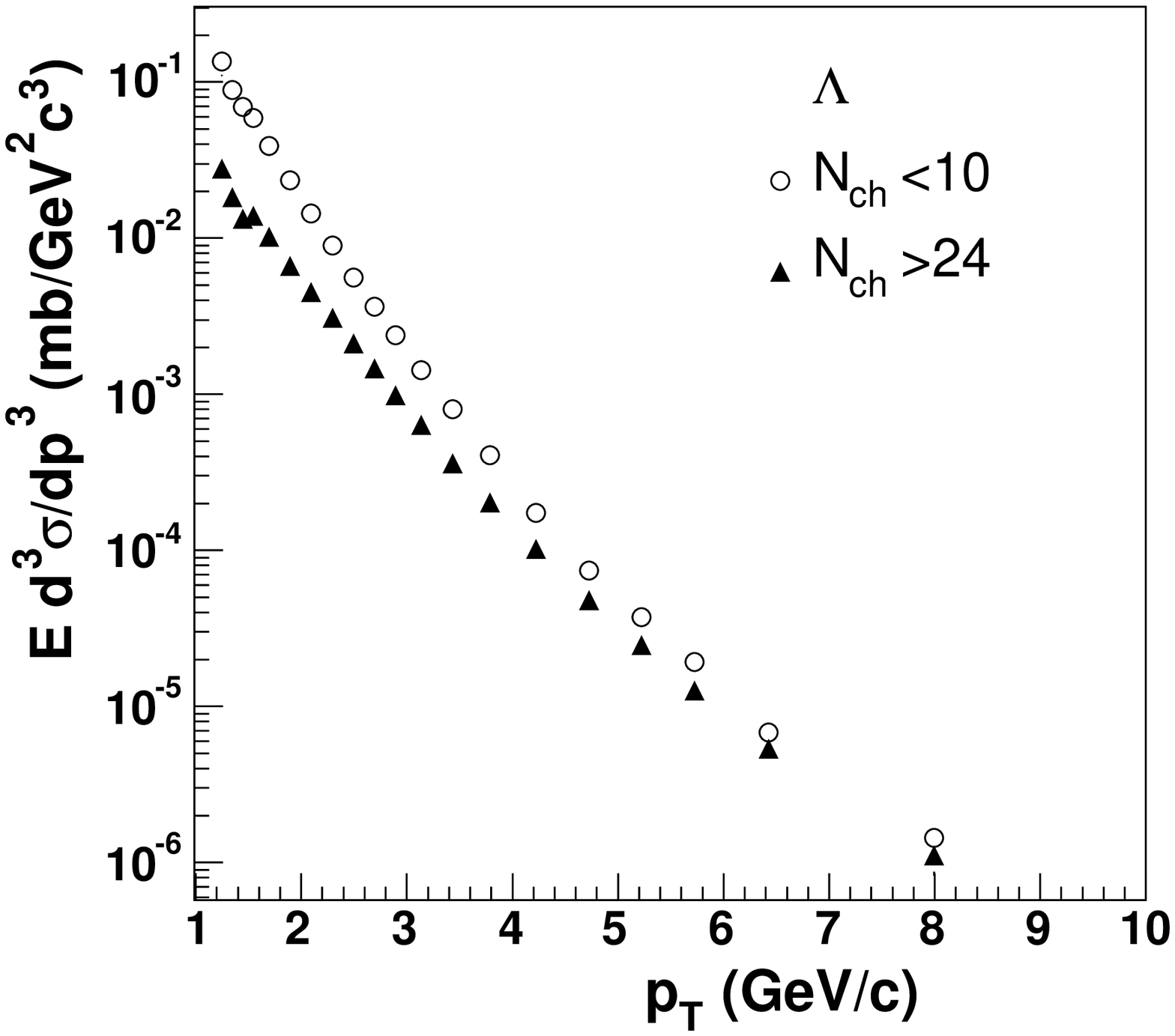,height=7cm}}
{\psfig{file=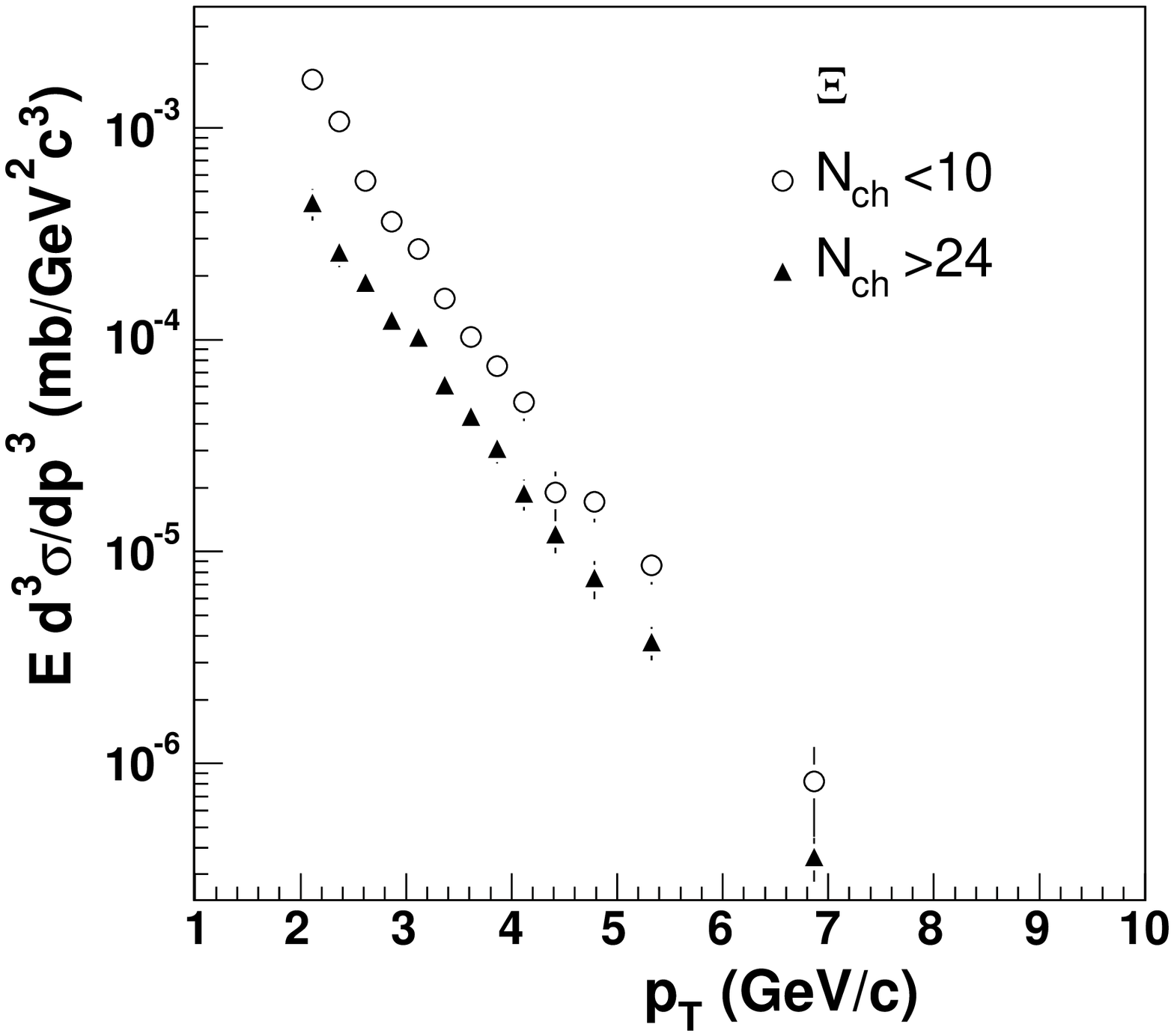,height=7cm}}
\caption{The $p_T$ differential cross sections for two charged-particle multiplicity
  	regions, $N_{ch}< 10$ and $N_{ch}>24$.  Distributions for
  	$\Lambda^0$ are shown on the left while distributions for
  	$\Xi^-$ are shown on the right.}
\label{f:c}
\end{figure*}

We thank the Fermilab staff and the technical staffs of the 
participating institutions for their vital contributions. This work 
was supported by the U.S. Department of Energy and National Science 
Foundation; the Italian Istituto Nazionale di Fisica Nucleare; the 
Ministry of Education, Culture, Sports, Science and Technology of 
Japan; the Natural Sciences and Engineering Research Council of 
Canada; the National Science Council of the Republic of China; the 
Swiss National Science Foundation; the A.P. Sloan Foundation; the 
Bundesministerium f\"ur Bildung und Forschung, Germany; the World 
Class University Program, the National Research Foundation of Korea; 
the Science and Technology Facilities Council and the Royal Society, 
UK; the Institut National de Physique Nucleaire et Physique des 
Particules/CNRS; the Russian Foundation for Basic Research; the 
Ministerio de Ciencia e Innovaci\'{o}n, and Programa Consolider-Ingenio 
2010, Spain; the Slovak R\&D Agency; and the Academy of Finland.


\begin{thebibliography}{99}
\bibitem{fn:a} 
 In the CDF coordinate system, $\theta$ and $\phi$ are the polar and azimuthal
 angles of a track, respectively, defined with respect to the proton beam
 direction, $z$. The pseudorapidity $\eta$ is defined as $-\ln [\tan(\theta/2)]$.
 The transverse-momentum of a particle is $p_T = p\sin{\theta}$. The rapidity 
 is defined as $y = 0.5 \ln [ (E+p_{z})/(E-p_{z}) ]$, where $E$ and $p_z$ are the 
 energy and longitudinal momentum of the particles associated with the track.
\bibitem{fn:b}
 The impact parameter $d_0$ is the distance 
 of closest approach of a track and the primary vertex in the $r-\phi$ plane.
\bibitem{fn:bb}
The number of $p_T$ intervals for data is dictated by statistics such that 
the fits to the invariant mass distributions are stable. 
In the acceptance calculation, the number of $p_T$ points is chosen such that 
a smooth acceptance curve as a function of $p_T$ can be obtained. The statistical 
uncertainties of the acceptance values are less than a few percent.
\bibitem{c:l} 
The total cross section corresponding to the minimum-bias trigger is estimated 
to be $45\pm 8$ mb. The elastic ($17\pm4$ mb~\cite{c:o}), single diffractive SD 
($12$ mb), and half of the double diffractive DD ($4$ mb) cross sections are 
subtracted from the total $p\overline{p}$ cross section ($78\pm6$ mb~\cite{c:o}) 
to give this estimate.  The SD and DD cross sections are estimated using {\sc pythia}.
A simulation study shows that the minimum-bias trigger is sensitive to $\sim100$\% 
of inelastic events which are not SD or DD and $\sim50$\% of DD events. 
A 100\% uncertainty is assigned to the DD contribution.
\end{thebibliography}
\end{document}